%% file: main.tex
\documentclass[12pt]{article}
\input{preamble/preamble}

\input{preamble/preamble_acronyms}
\input{preamble/preamble_math}

\input{preamble/preamble_tikz}

\title{%
\textbf{Discussion of ``Fast Approximate Inference for Arbitrarily Large Semiparametric
Regression Models via Message Passing''}
}

\author{
Dustin Tran \\
Columbia University \\
\\
David M.~Blei \\
Columbia University \\
}

\begin{document}

\maketitle
\bigskip


We commend \citet{wand2017fast} for an excellent description of
\gls{MP} and for developing it to infer large semiparametric
regression models.  We agree with the author in fully embracing the
modular nature of \acrlong{MP}, where one can define ``fragments''
that enable us to compose localized algorithms.  We believe this
perspective can aid in the development of new algorithms for automated
inference.

\parhead{Automated inference.}  The promise of automated algorithms is
that modeling and inference can be separated. A user can construct
large, complicated models in accordance with the assumptions he or she
is willing to make about their data. Then the user can use generic
inference algorithms as a computational backend in a ``probabilistic
programming language,'' i.e., a language for specifying generative
probability models.

With probabilistic programming, the user no longer has to write their
own algorithms, which may require tedious model-specific derivations
and implementations. In the same spirit, the user no longer has to
bottleneck their modeling choices in order to fit the requirements of
an existing model-specific algorithm.  Automated inference enables
probabilistic programming systems, such as
Stan~\citep{carpenter2016stan}, through methods like
\gls{ADVI}~\citep{kucukelbir2016automatic} and
\gls{NUTS}~\citep{hoffman2014nuts}.

Though they aim to apply to a large class of models, automated
inference algorithms typically need to incorporate modeling structure
in order to remain practical. For example, Stan assumes that one can
at least take gradients of a model's joint density. (Contrast this
with other languages which assume one can only sample from the model.)
However, more structure is often necessary: \gls{ADVI} and \gls{NUTS}
are not fast enough by themselves to infer very large models, such as
hierarchical models with many groups.

We believe \gls{MP} and Wand's work could offer fruitful avenues for
expanding the frontiers of automated inference. From our perspective,
a core principle underlying \gls{MP} is to leverage structure when it
is available---in particular, statistical properties in the model---
which provides useful computational properties.  In \gls{MP}, two
examples are conditional independence and conditional conjugacy.

\parhead{From conditional independence to distributed computation.}
As \citet{wand2017fast} indicates, a crucial advantage of message
passing is that it modularizes inference; the computation can be
performed separately over conditionally independent posterior
factors. By definition, conditional independence separates a posterior
factor from the rest of the model, which enables \gls{MP} to define a
series of iterative updates. These updates can be run asynchronously
and in a distributed environment.

\begin{figure}[tb]
  \centering
  \input{img/hierarchical_model.tex}
\caption{%
A hierarchical model, with latent variables $\alpha_k$ defined locally
per group and latent variables $\phi$ defined globally to be shared across groups.
}
\label{fig:hierarchical_model}
\end{figure}
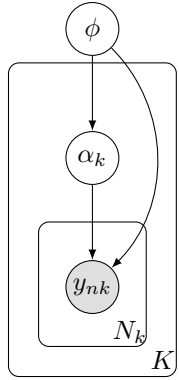

We are motivated by hierarchical models, which substantially benefit
from this property. Formally, let $y_{nk}$ be the $n^{\rm th}$ data
point in group $k$, with a total of $N_k$ data points in group $k$ and
$K$ many groups. We model the data using local latent variables
$\alpha_k$ associated to a group $k$, and using global latent
variables $\phi$ which are shared across groups. The model is depicted
in \Cref{fig:hierarchical_model}.

The posterior distribution of local variables $\alpha_k$ and global
variables $\phi$ is
\begin{equation*}
p(\alpha,\phi\g\mby) \propto
p(\phi\g\mby) \prod_{k=1}^K
\Big[ p(\alpha_k\g \beta) \prod_{n=1}^{N_K} p(y_{nk}\g\alpha_k,\phi) \Big].
\end{equation*}
The benefit of distributed updates over the independent factors is
immediate. For example, suppose the data consists of 1,000 data points
per group (with 5,000 groups); we model it with 2 latent variables per
group and 20 global latent variables.  Passing messages, or
inferential updates, in parallel provides an attractive approach to
handling all 10,020 latent dimensions. (In contrast, consider a
sequential algorithm that requires taking 10,019 steps for all other
variables before repeating an update of the first.)

While this approach to leveraging conditional independence is
straightforward from the message passing perspective, it is not
necessarily immediate from other perspectives.  For example, the
statistics literature has only recently come to similar ideas,
motivated by scaling up Markov chain Monte Carlo using divide and
conquer strategies~\citep{huang2005sampling,wang2013parallelizing}.
These first analyze data locally over a partition of the joint
density, and second aggregate the local inferences.  In our work in
\citet{gelman2014expectation}, we arrive at the continuation of this
idea. Like message passing, the process is iterated, so that local
information propagates to global information and global information
propagates to local information. In doing so, we obtain a scalable
approach to Monte Carlo inference, both from a top-down view which
deals with fitting statistical models to large data sets and from a
bottom-up view which deals with combining information across local
sources of data and models.

\parhead{From conditional conjugacy to exact iterative updates.}
Another important element of message passing algorithms is conditional
conjugacy, which lets us easily calculate the exact distribution for a
posterior factor conditional on other latent variables. This enables
analytically tractable messages (c.f., Equations (7)-(8) of
\citet{wand2017fast}).

Consider the same hierarchical model discussed above, and set
\begin{align*}
p(y_k,\alpha_k\g \phi)
&= h(y_k, \alpha_k) \exp\{\phi^\top t(y_k, \alpha_k) - a(\phi)\},
\\
p(\phi)
&= h(\phi) \exp\{\eta^{(0) \top} t(\phi) - a(\eta_0)\}
.
\end{align*}
The local factor $p(y_k,\alpha_k\g\phi)$ has sufficient statistics
$t(y_k,\alpha_k)$ and natural parameters given by the global latent
variable $\phi$.  The global factor $p(\phi)$ has sufficient
statistics $t(\phi) = (\phi, -a(\phi))$, and with fixed
hyperparameters $\eta^{(0)}$, which has two components: $\eta^{(0)} =
(\eta^{(0)}_1,\eta^{(0)}_2)$.

This exponential family structure implies that, conditionally, the
posterior factors are also in the same exponential families
as the prior factors~\citep{diaconis1979conjugate},
\begin{align*}
p(\phi\g\mby,\alpha)
&= h(\phi) \exp\{\eta(\mby,\alpha)^\top t(\phi) - a(\mby,\alpha)\},
\\
p(\alpha_k\g y_k, \phi)
&= h(\alpha_k) \exp\{\eta(y_k, \phi)^\top t(\alpha_k) - a(y_k, \phi)\}
.
\end{align*}
The global factor's natural parameter is $\eta(\mby,\alpha) =
(\eta^{(0)}_1 + \sum_{k=1}^K t(y_k, \alpha_k), \eta^{(0)}_2 + \sum_{k=1}^K N_k)$.

With this statistical property at play---namely that conjugacy gives
rise to tractable conditional posterior factors---we can derive
algorithms at a conditional level with exact iterative updates.  This
is assumed for most of the message passing of semiparametric models in
\citet{wand2017fast}.  Importantly, this is not necessarily a
limitation of the algorithm. It is a testament to leveraging model
structure: without access to tractable conditional posteriors,
additional approximations must be made.  \citet{wand2017fast} provides
an elegant way to separate out these nonconjugate pieces from the
conjugate pieces.

In statistics, the most well-known example which leverages
conditionally conjugate factors is the Gibbs sampling algorithm.  From
our own work, we apply the idea in order to access fast natural
gradients in variational inference, which accounts for the information
geometry of the parameter space~\citep{hoffman2013stochastic}.  In
other work, we demonstrate a collection of methods for gradient-based
marginal optimization~\citep{tran2016gradient}.  Assuming forms of
conjugacy in the model class arrives at the classic idea of
iteratively reweighted least squares as well as the EM algorithm. Such
structure in the model provides efficient algorithms---both
statistically and computationally---for their automated inference.

\parhead{Open Challenges and Future Directions.} \Acrlong{MP} is a
classic algorithm in the computer science literature, which is ripe
with interesting ideas for statistical inference. In particular,
\gls{MP} enables new advancements in the realm of automated inference,
where one can take advantage of statistical structure in the model.
\citet{wand2017fast} makes great steps following this direction.

With that said, important open challenges still exist in order to
realize this fusion.

First is about the design and implementation of probabilistic
programming languages. In order to implement \citet{wand2017fast}'s
message passing, the language must provide ways of identifying local
structure in a probabilistic program.  While that is enough to let
practitioners use \gls{MP}, a much larger challenge is to
then automate the process of detecting local structure.

Second is about the design and implementation of inference engines.
The inference must be extensible, so that users can not only employ
the algorithm in \citet{wand2017fast} but easily build on top of it.
Further, its infrastructure must be able to encompass a variety of
algorithms, so that users can incorporate \gls{MP} as one of many
tools in their toolbox.

Third, we think there are innovations to be made on taking the stance
of modularity to a further extreme. In principle, one can compose not
only localized message passing updates but compose localized inference
algorithms of any choice---whether it be exact inference, Monte Carlo,
or variational methods.  This modularity will enable new
experimentation with inference hybrids and can bridge the gap among
inference methods.

Finally, while we discuss \gls{MP} in the context of automation,
fully automatic algorithms are not possible. Associated to all
inference are statistical and computational
tradeoffs~\citep{jordan2013statistics}.  Thus we need algorithms along
the frontier, where a user can explicitly define a computational
budget and employ an algorithm achieving the best statistical
properties within that budget; or conversely, define desired
statistical properties and employ the fastest algorithm to achieve
them.  We think ideas in \gls{MP} will also help in developing some of
these algorithms.


\section*{References}
\renewcommand{\bibsection}{}
\bibliographystyle{apalike}
\bibliography{bib}

\end{document}

%% file: preamble/preamble.tex
\usepackage[T1]{fontenc}
\usepackage{tgtermes}

\usepackage{amssymb}
\newcommand{\mathbold}[1]{\ensuremath{\boldsymbol{\mathbf{#1}}}}

\usepackage{scalefnt,letltxmacro}
\LetLtxMacro{\oldtextsc}{\textsc}
\renewcommand{\textsc}[1]{\oldtextsc{\scalefont{1.10}#1}}
\usepackage[ttdefault=true]{AnonymousPro}

\usepackage{amsmath}

\usepackage[
  paper  = letterpaper,
  left   = 1.25in,
  right  = 1.25in,
  top    = 1.0in,
  bottom = 1.0in,
  ]{geometry}
\usepackage[english]{babel}
\usepackage[parfill]{parskip}
\usepackage{afterpage}
\usepackage{enumitem}
\usepackage{framed}
\usepackage{xspace}

\usepackage{lineno}

\usepackage{ragged2e}

\newcounter{parcount}

\DeclareRobustCommand{\parhead}[1]{\textbf{#1}~}
\usepackage[usenames,dvipsnames]{xcolor}
\definecolor{shadecolor}{gray}{0.9}

\usepackage{graphicx}
\usepackage[labelfont=bf]{caption}
\usepackage[format=hang]{subcaption}

\usepackage{booktabs, array}

\usepackage[algoruled]{algorithm2e}
\usepackage{listings}
\usepackage{fancyvrb}
\fvset{fontsize=\small}

\usepackage{natbib}
\usepackage[colorlinks,linktoc=all]{hyperref}
\usepackage[all]{hypcap}
\hypersetup{citecolor=MidnightBlue}
\hypersetup{linkcolor=MidnightBlue}
\hypersetup{urlcolor=MidnightBlue}
\usepackage{cleveref}
\creflabelformat{equation}{#2\textup{#1}#3}  

\usepackage
[acronym,smallcaps,nowarn,section,nonumberlist]{glossaries}
\glsdisablehyper{}

\lstdefinestyle{mystyle}{
    commentstyle=\color{OliveGreen},
    numberstyle=\tiny\color{black!60},
    stringstyle=\color{BrickRed},
    basicstyle=\ttfamily\scriptsize,
    breakatwhitespace=false,
    breaklines=true,
    captionpos=b,
    keepspaces=true,
    numbers=none,
    numbersep=5pt,
    showspaces=false,
    showstringspaces=false,
    showtabs=false,
    tabsize=2
}


%% file: preamble/preamble_acronyms.tex
\newacronym{MP}{mp}{message passing}
\newacronym{ADVI}{advi}{automatic differentiation variational inference}
\newacronym{NUTS}{nuts}{no U-turn sampler}

%% file: preamble/preamble_math.tex
\newcommand{\g}{\,|\,}

\newcommand{\mby}{\mathbold{y}}

%% file: preamble/preamble_tikz.tex
\usepackage{tikz}

\usetikzlibrary{bayesnet} 

\pgfdeclarelayer{edgelayer}
\pgfdeclarelayer{nodelayer}
\pgfsetlayers{edgelayer,nodelayer,main}

\definecolor{hexcolor0xbfbfbf}{rgb}{0.749,0.749,0.749}

\tikzset{>=latex}
\tikzstyle{none}   = [inner sep=0pt]
\tikzstyle{line}   = [ -, thick, shorten <=1pt, shorten >=1pt ]
\tikzstyle{arrow}  = [ ->, thick, shorten <=1pt, shorten >=1pt ]
\tikzstyle{ardash} = [ dashed, ->, thick, shorten <=1pt, shorten >=1pt ]

\tikzstyle{empty}=[circle,opacity=0.0,text opacity=1.0,inner sep=0pt]
\tikzstyle{box}=[rectangle,fill=White,draw=Black]
\tikzstyle{filled}=[circle,thick,fill=hexcolor0xbfbfbf,draw=Black]
\tikzstyle{hollow}=[circle,thick,fill=White,draw=Black]
\tikzstyle{param}=[rectangle,fill=Black,draw=Black,inner sep=0pt,minimum width=4pt,minimum height=4pt]
\tikzstyle{paramhollow}=[rectangle,thick,fill=White,draw=Black,inner sep=0pt,minimum
width=4pt,minimum height=4pt]

\usepackage{pgfplots}                               
\pgfplotsset{compat=newest}
\pgfplotsset{plot coordinates/math parser=false}
\newlength\figureheight
\newlength\figurewidth
\newlength\figureheightsmall
\newlength\figurewidthsmall
\definecolor{POSTcolor}{rgb}{0.48, 0.20, 0.58} 
\definecolor{Qcolor}{rgb}{0.00, 0.53, 0.22} 

%% file: img/hierarchical_model.tex
\begin{tikzpicture}[x=1.7cm,y=1.8cm]

  \node[latent]               (phi)      {$\phi$} ;
  \node[latent, below=1.0cm of phi] (alpha)    {$\alpha_k$} ;
  \node[obs, below=1.0cm of alpha]      (y)        {$y_{nk}$} ;

  \edge{phi}{alpha};
  \draw[->] (phi) to[out=-45,in=45] (y);
  \edge{alpha}{y};

  \plate[inner sep=0.75cm, yshift=0.15cm,
    label={[xshift=-14pt,yshift=14pt]south east:$K$}] {plate1} {
    (y)(alpha)
  } {};
  \plate[inner sep=0.35cm, yshift=0.15cm,
    label={[xshift=-17pt,yshift=14pt]south east:$N_k$}] {plate1} {
    (y)
  } {};

\end{tikzpicture}